
%
%
\magnification=\magstep1
\overfullrule=0pt
\setbox0=\hbox{{\cal W}}
\def\ww{{\cal W}\hskip-\wd0\hskip 0.2 true pt{\cal W}}
\def\w{{\cal W}}
\def\n{{\cal N}}
\def\S{{\cal S}}
\def\ob{{\bigg (}}
\def\cb{{\bigg )}}
\def\lb{\lbrack}
\def\rb{\rbrack}
\def\de{\partial}
\def\tr{{\rm tr}}
\def\q#1{\lb#1\rb}
\def\mn{\medskip\smallskip\noindent}
\def\sn{\smallskip\noindent}
\def\bn{\bigskip\noindent}
\font\extra=cmss10 scaled \magstep0 \font\extras=cmss10 scaled 750

\setbox1 = \hbox{{{\extra R}}}
\setbox2 = \hbox{{{\extra I}}}
\setbox3 = \hbox{{{\extra C}}}

\def\C{{{\extra C}}\hskip-\wd3\hskip2.5 true pt{{\extra I}}\hskip-\wd2
\hskip-2.5 true pt\hskip\wd3}
\def\Complex{\hbox{{\extra\C}}}
\def\One{{{\extra 1}}\hskip-\wd3\hskip0.5 true pt{{\extra 1}}\hskip-\wd2
\hskip-0.5 true pt\hskip\wd3}
\def\id{\hbox{{\extra\One}}}
\setbox4=\hbox{{{\extra Z}}}
\setbox5=\hbox{{{\extras Z}}}
\setbox6=\hbox{{{\extras z}}}
\def\Z{{{\extra Z}}\hskip-\wd4\hskip 2.5 true pt{{\extra Z}}}
\def\z{{{\extras Z}}\hskip-\wd5\hskip 2 true pt{{\extras Z}}}
\def\zs{{{\extras z}}\hskip-\wd6\hskip 1.7 true pt{{\extras z}}}
\def\Zed{\hbox{{\extra\Z}}}
\def\zed{\hbox{{\extras\z}}}
\def\zeds{\hbox{{\extras\zs}}}
\def\hwv{\mid \!h,w \rangle\, }
\def\ahwv{\langle h,w \! \mid\, }

\def\bpz{1}
\def\bouwschou{2}
\def\nahm{3}
\def\nam{4}
\def\kau{5}
\def\blm{6}
\def\rva{7}
\def\wirrep{8}
\def\mfl{9}
\def\wowoA{10}
\def\wowoB{11}
\def\hokimA{12}
\def\zam{13}
\def\kausch{14}
\def\dijkgraaf{15}
\def\romans{16}
\def\bbss{17}
\def\bai{18}
\def\deckmyn{19}
\def\wirsuperwrep{20}
\def\bil{21}
\def\watts{22}
\def\fateev{23}
\def\frenkel{24}
\def\hokimB{25}
\def\hokimC{26}
\def\commute{27}
\def\diplom{28}
\def\cap{29}
\def\ver{30}
\def\nahmpriv{31}
\def\felder{32}
\def\flohkau{33}
\def\baf{34}
\def\bal{35}
\def\blg{36}
\def\goddard{37}
\def\hornfeck{38}
\def\lykyanov{39}
\def\hgkpriv{40}
\def\supwir{41}
\def\zamzam{42}
\def\gehlen{43}
\def\cardy{44}
\def\schuetz{45}
\def\cardyA{46}
\def\gehlenA{47}
\def\henkel{48}
\def\baakeB{49}
\def\kohmoto{50}
\def\baakeA{51}
\def\alcaraz{52}
\def\gehlenunpub{53}
\def\gepner{54}
\def\saleur{55}
\def\duplantier{56}
\setbox9=\hbox{{\it k}}
\def\kk{{\it k}\hskip-\wd9\hskip 0.2 true pt{\it k}}
\def\A{{\cal A}}
\def\AC{{\cal C}}
\def\G{{\cal G}}
\def\D{{\cal D}}
\def\F{{\cal F}}
\def\I{{\cal I}}
\def\N{{\cal N}}
\def\FA{{\cal A}}
\def\FB{{\cal B}}
\def\Lie{{\cal L}}
\def\S{{\cal S}}
\def\mod{\phantom{l} {\rm mod} \phantom{l}}
\def\si{\sigma}
\def\Ga{\Gamma}
\def\la{\lambda}
\def\om{\omega}
\def\oh{{\textstyle{1 \over 2}}}
\font\HUGE=cmbx12 scaled \magstep4
\font\Huge=cmbx10 scaled \magstep4
\font\Large=cmr12 scaled \magstep3

\font\large=cmr17 scaled \magstep0
%
%
\nopagenumbers
\pageno = 0
\centerline{\HUGE Universit\"at Bonn}
\vskip 10pt
\centerline{\Huge Physikalisches Institut}
\vskip 2cm
\centerline{\Large Automorphisms of W-Algebras}
\vskip 6pt
\centerline{\Large \phantom{g} and \phantom{g}}
\vskip 6pt
\centerline{\Large Extended Rational Conformal Field Theories}
\vskip 0.4cm
\centerline{by}
\vskip 0.4cm
\centerline{\large A.\ Honecker}
\vskip 1.5cm
\centerline{\bf Abstract}
\vskip 15pt
\noindent
Many extended conformal algebras with one generator in addition to the
Virasoro field as well as Casimir algebras have non-trivial outer
automorphisms which enables one to impose `twisted' boundary
conditions on the chiral fields. We study their effect on the highest weight
representations. We give formulae for the enlarged rational
conformal field theories in both series of $\w$-algebras with two
generators and conjecture a general formula for the additional models
in the minimal series of Casimir algebras. A third series of
$\w$-algebras with two generators which includes the spin three
algebra at $c=-2$ also has finitely many additional fields in the twisted
sector although the model itself is apparently not rational.
The additional fields in the twisted sector have applications in
statistical mechanics as we demonstrate for $\Zed_n$-quantum spin chains
with a particular type of boundary conditions.
\vfill
\settabs \+&  \hskip 110mm & \phantom{XXXXXXXXXXX} & \cr
\+ & Post address:                       & BONN-HE-92-37   & \cr
\+ & Nu{\ss}allee 12                     & hep-th/9211130  & \cr
\+ & W-5300 Bonn 1                       & Bonn University & \cr
\+ & Germany                             & November 1992   & \cr
\+ & e-mail:                             & ISSN-0172-8733  & \cr
\+ & unp06b@ibm.rhrz.uni-bonn.de         &  \              & \cr
\eject
\pageno=1
\footline{\hss\tenrm\folio\hss}
%
\leftline{\bf 1.\ Introduction}
\mn
Rational conformal field theories (RCFTs) have attracted much attention
after they were introduced in the seminal work of Belavin et al.\ $\q{\bpz}$.
While the classification of all RCFTs is still an open question there has been
some progress studying extended conformal algebras, so-called $\w$-algebras
(for a recent review see e.g.\ $\q{\bouwschou}$).
With the help of explicit formulae
for local chiral algebras $\q{\nahm} \q{\nam}$
many new $\w$-algebras with two and three generators were
constructed $\q{\kau} \q{\blm}$. The study of their highest weight
representations (HWRs) revealed new rational models
$\q{\rva - \wowoB}$.
By now, there is a good chance that all
rational models belonging to $\w$-algebras with one additional generator
are classified -- although there is no proof of this fact yet.
\sn
Already some time ago, Q.\ Ho-Kim and H.B.\ Zheng $\q{\hokimA}$ have noticed
that Zamolodchikov's $\w(2,3)$ $\q{\zam}$ along with other Casimir algebras
have non-trivial outer automorphisms and therefore admit `twisted' boundary
conditions. As we will show in this paper, this is valid also for
certain extended conformal algebras with two generators (one in addition to
the Virasoro field). We will show that these twists lead to additional
HWRs and thus enlarge the RCFTs. One could also take a different point
of view and project the $\w$-algebra onto the invariant subspace. This
is called `orbifolding'. We will say more about the precise connection
between these two approaches later on.
\mn
The outline of this paper is as follows:
In the next two subsections we will briefly summarize our approach
to $\w$-symmetry and introduce twisted boundary conditions. Section 2
focusses on the well known example $\w(2,3)$ and illustrates our methods.
In section 3 we discuss the three series of bosonic $\w(2,\delta)$-algebras
that admit twists. Section 4 contains a discussion of twists of
Casimir algebras. Finally, in section 5 we shall show that some of the
additional representations can indeed be realized in statistical
mechanics models. This shall be
demonstrated in the case of $\Zed_n$-spin quantum chains with a
special type of boundary conditions.
\bn
\leftline{\bf 1.1.\ Theorems about $\ww$-algebras}
\mn
Before starting the main issues, we would like to summarize
briefly the notions
which we will need in this paper. For precise definitions and explicit
formulae, however, we refer to $\q{\blm} \q{\wirrep}$.
\sn
Let $\F$ be a local chiral conformal field theory. On $\F$ there are
three important operations: A commutator, a normal ordered product $N$
and the usual derivative $\de$.
Any field $\phi$ in $\F$ can be written as
$$\phi(z) = \sum_{n-d(\phi) \in \zed} z^{n-d(\phi)} \phi_n.
\eqno({\rm 1.1.1})$$
$d(\phi)$ is called the `conformal dimension' of $\phi$ and $\phi_n$ the
`modes' of $\phi$.
The modes of the energy-momentum tensor $L$ in $\F$ are well known to satisfy
the Virasoro algebra (the explicit form is given in the first line of (2.1)).
If the commutator of a field $\phi$ with the Virasoro algebra yields only
the field $\phi$ itself, $\phi$ is called `primary'. If this holds
only for the $SU(1,1)$-subalgebra spanned by $L_{-1}$, $L_0$ and $L_1$,
the field $\phi$ is called `quasiprimary'.
\sn
There is a general formula for the commutator of two quasiprimary local
\footnote{${}^{1})$}{fields with dimension in ${\zed_{+} \over 2}$.}
chiral fields $\q{\nahm}$.
The Lie bracket structure in $\F$ is fixed by
universal polynomials $p_{ijk}$ depending exclusively
on the conformal dimensions,
and a few structure constants which are basically given
by the two- and three-point-functions. Since the usual normal ordered
product $N(\phi, \de^n \psi)$ of two quasiprimary fields $\phi$ and $\psi$ is
not quasiprimary any more, we use a quasiprimary
normal ordering prescription $\N(\phi, \de^n \psi)$ $\q{\nahm}$. The explicit
formula is somewhat lengthy (see e.g.\ $\q{\blm})$ but the basic result
is rather simple: From $N(\phi, \de^n \psi)$ all fields turning up in the
commutator of $\phi$ and $\psi$ have to be subtracted with
dimension-dependent factors. From any finite set of fields the operations
$\de$ and $\N$ will generate infinitely many fields. It is therefore convenient
to define `simple' fields which are non-composite and non-derivative.
The algebra generated by simple fields $\phi_1 \ldots \phi_n$ is called
a `$\w(d(\phi_1), \ldots , d(\phi_n))$'.
\medskip
In this paper we will only consider $\w$-algebras where the zero modes of
the simple fields commute and thus may be considered as the Cartan
subalgebra. For these $\w$-algebras a highest weight representation
may be defined $\q{\wirrep}$ via the existence
of a cyclic vector $\hwv$ which is an eigenvector of the zero modes of all
simple bosonic fields (i.e.\ the fields with integer dimension) and
satisfies
$$\phi_n \hwv = 0 \qquad \forall \phi, \forall n<0. \eqno({\rm 1.1.2})$$
For a $\w$-algebra with two generators we will denote the additional simple
field by $W$ and the corresponding eigenvalues of the energy-momentum tensor
$L$ and the field $W$ by $h$ and $w$:
$$L_0 \hwv = h \hwv \ , \qquad W_0 \hwv = w \hwv. \eqno({\rm 1.1.3})$$
In order to define correlation functions one introduces a linear form
$\ahwv$ dual to $\hwv$ $\q{\wirrep}$.
\mn
There are two main approaches to the explicit study of the HWRs of
$\w$-algebras. The first one is based on the fact that only those HWRs
are physically relevant which vanish identically under the application
of null fields (fields with zero two point
functions in the vacuum representation). Thus, writing down
states containing null fields and demanding that they
should vanish yields conditions on $\hwv$. We shall give an example for this
approach in section 2. The second approach is based on the observation
of R.\ Varnhagen $\q{\rva}$ that correlation functions in HWRs are not
automatically associative, even if the algebra is. Basically, one studies
special Jacobi identities in order to check whether the commutator
$\lb \phi_n, \psi_m \rb$ 
is represented by $\phi_n \psi_m -
\psi_m \phi_n$. For more details see $\q{\wirrep}$.
\bn
\leftline{\bf 1.2.\ Automorphisms of $\ww$-algebras and boundary conditions}
\mn
In this paper we shall be interested in non-trivial outer automorphisms
of $\w$-algebras and their effect on the HWRs. An automorphism $\rho$ of
a $\w$-algebra is a bijective map of the algebra that is compatible
with the Lie bracket structure and the normal ordering prescription.
$\rho$ is called an `outer' automorphism if it is not generated by
the $\w$-algebra itself.
Each such automorphism enables one to impose non-trivial boundary
conditions on the fields $\phi_j$ in the algebra:
$$\phi_j\left(e^{2 \pi i} z\right) = \rho\left(\phi_j\left( z \right)\right).
\eqno({\rm 1.2.1})$$
This type of boundary condition will be called a `twist'.
\mn
The free fermion $\psi$ with commutation relations
$$\lb \psi_m, \psi_n \rb_{+} = \delta_{n, -m}   \eqno({\rm 1.2.2})$$
provides us with a simple example for such an automorphism:
$\rho(\psi)=-\psi$. More generally, for any $\w$-algebra that
contains fermionic fields, there is an automorphism of the following
type:
$$\eqalign{
\rho(\phi_j) &= \phi_j \phantom{-}\qquad \forall \phi_j: d(\phi_j) \in
        \Zed \cr
\rho(\psi_j) &= -\psi_j \qquad \forall \psi_j: d(\psi_j) \in
        \Zed + {\textstyle {1 \over 2}}. \cr
}\eqno({\rm 1.2.3})$$
Note that the Lie bracket structure as well as the normal ordered product
respect the $\Zed_2$-grading of the $\w$-algebra into bosonic and
fermionic fields. This shows that (1.2.3) indeed is an automorphism
of the $\w$-algebra.
Since for any field with $\phi_j\left(e^{2 \pi i} z\right) = -\phi_j(z)$
the Laurent expansion reads
$$\phi_j(z)=
\sum_{n-d(\phi_j)\in\zed+{1\over 2}} z^{n-d(\phi_j)}\phi_{j,n}\,
\eqno({\rm 1.2.4})$$
this leads to the Ramond-sector of a fermionic $\w$-algebra.
For the algebra $\w(2,3)$ the reflection of the additional bosonic field
$W$ with dimension three
$\rho(W) = -W$ is an automorphism of the algebra and leads to
half-integral modes of the field $W$ $\q{\hokimA}$. In the next section
we will illustrate the effect of the boundary conditions in this well
known example.
\sn
For general bosonic $\w(2, \delta)$-algebras an outer automorphism of this kind
exists iff the self coupling constant vanishes.
We recall that there are exactly three series that admit an automorphism
$\rho$ with:
$$\rho(L) = L, \qquad \rho(W) = -W               \eqno({\rm 1.2.5})$$
and therefore admit half-integral modes for the additional bosonic field:
$$W(z)=\sum_{n-\delta \in \zed+{1\over 2}} z^{n-\delta}W_{n}.
\eqno({\rm 1.2.6})$$
The first of these series is related to Virasoro minimal models
$\q{\rva}$, a second so-called `parabolic' series
exists for $c=1-8 \delta$ $\q{\mfl}$
and a third series exists for $c=c_{1,k}$ $\q{\kausch}$. They will be
discussed in three subsequent subsections of section 3.
\medskip
It is well known that one can also project the $\w$-algebra
onto the invariant subspace and then study the representations of the
orbifold (for a detailed discussion of orbifolding see e.g.\
$\q{\dijkgraaf}$). However, there is
a close connection between both approaches. We prefer to study the original
algebra with twisted boundary conditions because it is no trivial question
to find out for a given $\w$-algebra to which algebra an orbifold construction
leads. Furthermore, the orbifolds tend to have more generators (with higher
dimensions) than the original $\w$-algebra and thus it is far more difficult
to study their representations explicitly.
\bn
\leftline{\bf 2.\ $\bf \ww(2,3)$: The method and results}
\mn
$\w(2,3)$ has been written down as early as 1985 by A.B.\ Zamolodchikov
$\q{\zam}$. Although this algebra is well known we shall use it to
illustrate our notations
and methods. In our notation $\w(2,3)$ is given by the following commutation
relations of the simple fields:
$$\eqalign{
\lb L_m, L_n \rb &= (n - m) L_{m+n} +{c \over 12} (n^3-n) \delta_{n,-m} \cr
\lb L_m, W_n \rb &= (n - 2m) W_{m+n} \cr
\lb W_m, W_n \rb &= C_{WW}^L p_{332}(m,n) L_{m+n}
                  + C_{WW}^\Lambda p_{334}(m,n) \Lambda_{m+n}
                  +  {c \over 3} {n + 2 \choose 5} \delta_{n,-m} \ , \cr}
 \eqno(2.1)$$
where
$$\Lambda = \n(L,L) = N(L,L) - {3 \over 10} \partial^2 L \eqno({\rm 2.2a})$$
$$C_{WW}^L = 2, \ \ \ \ \ C_{WW}^\Lambda = {32 \over 5 c + 22}
         \eqno({\rm 2.2b})$$
$$p_{334}(m,n) = {n-m \over 2} \ , \ \ \ \ \
p_{332}(m,n) = {n-m \over 60} ( 2 m^2 - m n + 2 n^2 - 8) \ .
  \eqno({\rm 2.2c})$$
\mn
The representation theory of $\w(2,3)$ is well studied
(see e.g.\ $\q{\romans}$): Coset constructions using $su(3)$
$\q{\bbss} \q{\bai}$ and even more general affine Lie algebras $\q{\deckmyn}$
are known in the literature.
Still, one can derive interesting new results for this algebra studying
null fields as we shall do now.
\mn
First, we consider the case $c={4 \over 5}$.
Here, one can calculate that the fields
$$ \eqalign{
\FA &:= \n(W,W) - {95 \over 117} \n(\Lambda, L)
                + {11 \over 6} \n(L, \partial^2 L), \cr
\FB &:= \n(W,\partial^2 L) - {315 \over 1196} \n(\n(W,L), L) \cr
}\eqno(\rm{ 2.3})$$
have vanishing two point functions and thus
are null fields. A quite effective set of null states is given by:
$$\eqalign{
\FA_0 \hwv &= -{1 \over 585} \ob (95 h - 7) (5 h - 2) h - 585 w^2
                             \cb  \hwv \cr
\FB_0 \hwv &= -{7 \over 1196} (15 h - 1) (3 h - 2) w \hwv \cr
\FB_{-1} \FB_1 \hwv &=
{49\over 836793360} \ob 1170 (225 h^2 - 160 h + 14) (90 h - 71) w^2 \cr
     &+ (80 h + 3) (27 h - 7)^2
     (5 h - 2)^2 h \cb  \hwv \cr
\FB_{-2} \FB_2 \hwv &=
{49 \over 209198340} \ob
      585 (20250 h^3 + 6075 h^2 - 2605 h - 5734) w^2 \cr
     &+ 2 ( 72900 h^4 + 610605 h^3 - 570491 h^2 + 38684
      h - 2184) (5 h - 2) h \cb \hwv. \cr
} \eqno(\rm{ 2.4})$$
Thus, for $c={4 \over 5}$, we necessarily have either
$w=0$ or $h \in \{ {1 \over 15}, {2 \over 3} \}$.
For $w=0$ we additionally have $h \in \{ 0, {2 \over 5} \}$ whereas
for the two other $h$-values (2.4) also fixes $w^2$. We conclude that
at most four HWRs are physically relevant for $c={4 \over 5}$
(see also table below).
\medskip
{}From (2.1) we explicitly see that $\w(2,3)$ is invariant under
$W \mapsto -W$ for generic value of $c$. Thus we can twist the field $W$
as described in section 1.2. Having done this, there will be no zero
mode of $W$ left,
such that we have to fix the eigenvalue of $L_0$ only and
a single condition might be sufficient (compared to (2.4))
\sn
Let us now return to the case $c={4 \over 5}$. Like for
Super-$\w$-algebras $\q{\wirsuperwrep}$, it is not possible to evaluate
the normal ordered product of two twisted fields using the standard formula
due to the non-local effect of the boundary conditions. Thus, the part
$\N(W,W)$ of $\FA$ cannot be evaluated easily and we only consider the
field $\FB$. In order to derive an equation for the eigenvalue of $L_0$
we have to consider the product of two modes of $\FB$ with vanishing
total grade. One evaluates e.g.:
$$\eqalign{
\FB_{-{1 \over 2}} \FB_{1 \over 2} \hwv
  &= p_1(h) (40 h - 1)^2 (8 h - 1) \hwv \cr
\FB_{-{3 \over 2}} \FB_{3 \over 2} \hwv
  &= p_2(h) (40 h - 1) (8 h - 1) \hwv \cr
} \eqno({\rm 2.5})$$
with $p_1$ and $p_2$ two coprime polynomials in $h$.
Thus, we find physically relevant representations in the twisted sector
of $\w(2,3)$ at $c={4 \over 5}$
only for $h \in \{ {1 \over 40}, {1 \over 8} \}$.
\bigskip
In principle, one can study all representations of $\w(2,3)$ examining null
fields. Among the fields with dimension not larger than seven,
we find at least two null fields for
$c \in \{ {4 \over 5}, -2, -23, -{114 \over 7} \}$.
It should be clear to the reader that the restriction to these cases is a
technical one rather than a principal one.
Going through the same steps as before we find
rational theories for $c=-23$ and $c=-{114 \over 7}$ in addition to
$c={4 \over 5}$. The values of $h$ for these theories are contained in the
following table:
\mn
\centerline{
\vrule \hskip 1pt
\vbox{ \offinterlineskip
\def\tablespace{height2pt&\omit&&\omit&&\omit&&\omit&&\omit&&\omit&&\omit&\cr}
\def\tablerule{ \tablespace
                \noalign{\hrule}
                \tablespace        }
\hrule
\halign{&\vrule#&
  \strut\hskip 4pt\hfil#\hfil\hskip 4pt\cr
height4pt& \multispan{13}  & \cr
& \multispan{13} \hfil $\w(2,3)$ \hfil &\cr
height4pt& \multispan{13}  & \cr
\noalign{\hrule}
height3pt& \multispan3 && \multispan3 && \multispan3 && \omit & \cr
& \multispan3 \hfil $c=c_{4,5}^{\A_2}={4 \over 5}$ \hfil
   && \multispan3 \hfil $c=c_{3,8}^{\A_2}=-23$ \hfil
         && \multispan3 \hfil $c=c_{3,7}^{\A_2}=-{114 \over 7}$ \hfil &&
                     $c=-2$ &\cr
height3pt& \multispan3 && \multispan3 && \multispan3 && \omit & \cr
\noalign{\hrule}
\tablespace
& {\it untwisted} && {\it twisted}   && {\it untwisted}  && {\it twisted}
   && {\it untwisted} && {\it twisted}    &&  {\it twisted} &\cr \tablerule
&  $0$            && ${1 \over 8}$   && $0$              && $-{31 \over 32}$
   && $0$             && $-{5 \over 8}$   && $-{3 \over 32}$ &\cr \tablespace
&  ${2 \over 3}$  && ${1 \over 40}$  && $-{1\over 2}$    && $-{23 \over 32}$
   && $-{3 \over 7}$  && $-{19 \over 56}$ && ${5 \over 32}$ &\cr \tablespace
&  ${2 \over 5}$  && \omit           && $-{3 \over 4}$   && $-{15 \over 16}$
   && $-{4 \over 7}$  && $-{39 \over 56}$ && \omit          &\cr \tablespace
&  ${1 \over 15}$ && \omit           && $-{7 \over 8}$   && \omit
    && $-{5 \over 7}$  && \omit           && \omit          &\cr \tablespace
&  \omit          && \omit           && $-1$             && \omit
   && \omit           && \omit            && \omit          &\cr \tablespace
}
\hrule}\hskip 1pt \vrule
}
\mn
We have also included the $h$-values of the twisted sector for $c=-2$ in
this table although in the untwisted sector we were not able to exclude
any pair $(h,w)$ satisfying the following relation:
$$w^2 = {2 \over 27} (8 h + 1) h^2.   \eqno({\rm 2.6})$$
In fact, $c=-2$ does not belong to the minimal series of $\w(2,3)$
given by
$$\eqalign{
c_{p,q}^{\A_2} &= 2  \ob 1-12{(p-q)^2 \over p q } \cb \cr
h_{p,q;r_1,s_1,r_2,s_2}^{\A_2} &=
 {3 \ob q (r_1+r_2)-p (s_1+s_2)\cb^2
   +\ob q (r_1-r_2)-p (s_1-s_2)\cb^2 \over 12 p q}
   +{c_{p,q}^{\A_2}-2 \over 24} \cr
} \eqno({\rm 2.7}) $$
with coprime $q>p$, $p>2$ as well as $r_1 + r_2 < p$ and $s_1 + s_2 < q$.
The parametrization of $c$ in the minimal series was well known
(see e.g.\ $\q{\bil} \q{\watts}$) whereas the values of $h$ were known
only for the unitary case $q=p+1$ (see e.g.\ $\q{\fateev}$). With the
above data we guessed that the formula for the $h$-values
can be extended to the non-unitary case simply by replacing $p+1$ with $q$.
Recently, (2.7) has been proved using quantized
Drinfeld-Sokolov reduction $\q{\frenkel}$. Note that
(2.7) correctly reproduces the representations of the untwisted sector in
the table above.
\mn
In the case of twisted $\w(2,3)$ up to now only a formula for the minimal
unitary series was known $\q{\hokimB} \q{\hokimC}$. Rewriting this
result as in (2.7), i.e.\ identifying the contribution of the central
charge to the conformal dimensions and replacing $p+1$ by $q$, we
conjecture that the conformal dimensions in the complete minimal
series of $\w(2,3)$ are given by
$$h_{p,q;r,s}^{\A_2^{(2)}} =
 \Omega^2 {(s p - r q)^2 \over 2 p q }
 +{c_{p,q}^{\A_2}-2 \over 24}
 +\tilde{h}.
\eqno({\rm 2.8}) $$
In the unitary minimal series we have $\Omega^2={1 \over 2}$
and the dimension
of the twisted field is $\tilde{h} = {1 \over 16}$ $\q{\hokimB} \q{\hokimC}$.
Indeed, with this
choice the first values for $r=1$ and $s \ge 1$ correctly reproduce the
$h$-values for $c={4 \over 5}$ and $c=-23$ in the table above. However, for
$c=-{114 \over 7}$ we need $\Omega^2=1$ and $\tilde{h} = {1 \over 24}$.
Note that these two cases differ in $p+q$ being odd or even, respectively,
though we do not know whether this already covers all possible cases.
\mn
Although we can parametrize $-2=c_{2,3}^{\A_2}$, this model
is not minimal because the condition
$p>2$ is violated. In fact, this model belongs to the series of non-minimal
models discussed in section 3.3.
\bn
\leftline{\bf 3.\ Explicit results}
\mn
In the following three subsections we will give formulae that correctly
reproduce the HWRs in the twisted sector of all bosonic
$\w(2, \delta)$-algebras with $3 < \delta \le 8$.
The explicit calculations
were performed on a computer using a special purpose program
$\q{\commute}$ supported by the computer algebra system REDUCE. In all
cases with $\delta > 3$ both methods mentioned in section 1 lead to
identical results. The values of $h$ and $w$, however, will not be
presented explicitly here. The interested reader may look them up in
$\q{\diplom}$.
\sn
Although we expect our formulae to be valid
for all members of the three series to be discussed below,
we will not prove this rigorously. With respect to the classification
problem we should mention that there is good reason
to believe that these three series include all $\w(2, \delta)$-algebras with
vanishing self coupling constant and $\delta > 3$ but there is no proof of
this fact yet.
\bn
\leftline{\bf 3.1.\ $\ww$-algebras related to Virasoro minimal models}
\mn
As already pointed out in $\q{\blm}$ many $\w(2, \delta)$-algebras are related
to Virasoro-minimal models. The central charge and conformal dimensions of
the primary fields in Virasoro-minimal models are given for any coprime $p$,
$q$ by the following expressions:
$$c_{p,q} = 1 - 6{(p-q)^2 \over pq}, \eqno({\rm 3.1.1a})$$
$$h_{p,q;r,s} = {(pr - qs)^2 - (p - q)^2 \over 4pq}\,,\ \
1 \leq r \leq q-1\,,\ \, 1 \leq s \leq p-1\,. \eqno({\rm 3.1.1b})$$
The representation theory of the corresponding
$\w$-algebras can be reduced
to the representation theory of the Virasoro algebra.
This has been noticed and worked out for fermionic $\w(2, \delta)$-algebras
in $\q{\rva}$. The formulae for the untwisted sector of bosonic $\w$-algebras
have been presented in $\q{\wirrep}$.
\sn
In the ADE-classification of Cappelli et al.\ $\q{\cap}$ there are
non-diagonal partition functions in terms of Virasoro-minimal models.
Now one can extend the symmetry algebra by those fields where the
corresponding characters turn up in the same summand as the character
for $h=0$ which corresponds to the field $L$. Additional arguments that this
procedure of extending the symmetry algebra does indeed yield a closed algebra
can be inferred from fusion rule considerations $\q{\blm}$. In this case, the
characters $\chi^W$ of the $\w$-algebra can be written as finite sums of
characters $\chi$ of the Virasoro algebra such that the characters $\chi^W$
diagonalize the partition function. For these models, the central charge
and the conformal dimensions can be parametrized according to (3.1.1).
The modular transformations of the characters $\chi^W$ and -- via the Verlinde
formula $\q{\ver}$ -- the fusion rules can be derived from those of the
corresponding Virasoro-minimal models. For fermionic $\w(2,\delta)$-algebras
this has been worked out in detail in $\q{\rva}$.
\medskip
The characters $\chi$ and $\chi^W$ are defined as a trace over the
representation module $V$ with formal powers in $q$:
$$\chi := \tr_V \ob q^{(L_0 - {c \over 24})} \cb . \eqno({\rm 3.1.2})$$
If we can write a character $\chi^W$ as a sum of Virasoro characters,
(3.1.2) shows that the value of $h$ for the HWR of the $\w$-algebra is
the smallest of the corresponding Virasoro HWRs.
\medskip
The case we shall be interested in here is described by the proposition in
section 4 of $\q{\blm}$ and is closely related to the
$(A_{q-1}, D_{2 n})$-series of modular invariant partition functions.
This series includes exactly those bosonic $\w(2, \delta)$-algebras which
are related to Virasoro minimal models and have vanishing self coupling
constant. For these $\w(2,\delta)$-algebras
$\delta = (q-2) (n-1)$ and $c = c_{(4 n - 2), q}$ holds. In
particular, the effective central charge $\tilde{c} = c - 24_{h_{min}}$ always
satisfies $\tilde{c} < 1$.
\sn
We claim that the characters $\chi^W$ are given by:
$$\eqalign{
\chi^W_{i,j} &= \chi^{}_{i,j} + \chi^{}_{q-i,j} \ , \ \ \
1 \leq i \leq {q \over 2} \ , \ \ \ 1 \leq j < {p \over 2} \ , \ \ \
i, j \in \Zed , \ j \in \I \cr
\chi^W_{i,{p \over 2}} &= \chi^{}_{i,{p \over 2}} \phantom{XXXXl} , \ \ \
1 \leq i \leq {q \over 2} \ , \ \ \ i \in \Zed , \ {p \over 2} \in \I. \cr
}\eqno({\rm 3.1.3})$$
Consequently, the values of $h$ are given by:
$$h^W_{i,j} = h_{i,j} \ , \ \ \ 1 \leq i \leq {q \over 2} \ , \ \ \
1 \leq j \leq {p \over 2} \ , \ \ \ i \in \Zed, \ j \in \I.
 \eqno({\rm 3.1.4})$$
For fermionic algebras we obtain for $\I=2 \Zed + 1$ the
Neveu-Schwarz-sector and for $\I=2 \Zed$ the Ramond-sector $\q{\rva}$.
For bosonic algebras $\I=2 \Zed + 1$ yields the untwisted sector of the
algebra which has already been discussed in $\q{\wirrep}$. Our explicit
results for bosonic $\w(2, \delta)$-algebras
show that with $\I = 2 \Zed$ one obtains the twisted sectors of these algebras.
Thus, adding a twisted sector, the bosonic algebras become similiar to
the fermionic ones.
\sn
Let us give a further argument that (3.1.3) gives the correct characters
also in the twisted sector of the bosonic algebras. If we extend the symmetry
algebra by a field $W$ we can use the modes of this field for a mapping
of two different Virasoro representation modules $V_1$ and $V_2$.
Thus, the representation module for the $\w$-algebra is given by
$V_1 \oplus V_2$ and the corresponding characters have to be added.
Now, the difference of the $h$-values of the two Virasoro HWRs has to
equal some mode of the field $W$ and therefore has to be integral in the
untwisted sector and half-integral in the twisted sector. Generically, the
only characters meeting this requirement are those given by (3.1.3).
\medskip
However, the action of the modular group on the characters of bosonic algebras
is quite different from the fermionic case because one has to impose
different boundary conditions in order to obtain a modular invariant
partition function.
It is well known that the untwisted sector of a bosonic $\w(2, \delta)$-algebra
is invariant under the full modular group generated by $S$ and $T$.
The corresponding partition function is non-diagonal in terms of
Virasoro-characters. The twisted sector is in contrast only invariant
under $T^2$ and $TST$ (in the fermionic case $TST$ intertwines
Neveu-Schwarz- and Ramond-sector).
If we want to act with the full modular group on it
we have to add further characters $\tilde{\chi}^W$:
$$
\tilde{\chi}^W_{i,j} = \chi^{}_{i,j} - \chi^{}_{q-i,j} \ , \ \ \
1 \leq i \leq {q \over 2} \ , \ \ \ 1 \leq j < {p \over 2} \ , \ \ \
i, j \in \Zed , \ j \in \I.
\eqno({\rm 3.1.5})$$
Now we can use all characters $\chi^W$ and $\tilde{\chi}^W$ to
write down a new diagonal modular invariant partition function.
If one expresses this `new' partition function in terms of Virasoro
characters one reobtains the original diagonal partition function,
i.e.\ the $(A_{q-1},A_{4 n - 3})$ modular invariant partition function
which is diagonal in terms of Virasoro characters
\footnote{${}^{1})$}{For $\w(2,3)$ at $c={4 \over 5}$ this has
already been pointed out in $\q{\hokimA}$.
}.
Still, the additional fields with dimensions given by (3.1.4) are
physically relevant as we shall show for $\w(2,3)$ at $c={4 \over 5}$
in section 5.
\bigskip
The series of $\w$-algebras related to Virasoro minimal models provides
us with a good example to understand the precise connection between
boundary conditions and orbifold constructions. It is a general feature
of $\w$-algebras with a $\Zed_2$-automorphism $\rho^2 = \id$ that
one has two partition functions, one where only the characters $\chi^W$
of the untwisted sector enter, and one where the characters $\chi^W$
and $\tilde{\chi}^W$ of both sectors enter. The latter can be identified
with the partition function $Z$ of the orbifold
\footnote{${}^{2})$}{We simplify notation by absorbing multiplicities
of characters into the index set.}:
$$\eqalign{
Z &= \sum_{k: \ {\rm untwisted}} (\chi_k^W)^{*} \chi_k^W
                          + (\tilde{\chi}_k^W)^{*} \tilde{\chi}_k^W
   + \sum_{k: \ {\rm twisted}} (\chi_k^W)^{*} \chi_k^W
                          + (\tilde{\chi}_k^W)^{*} \tilde{\chi}_k^W \cr
&= 2 \sum_{k: \ {\rm untwisted} \atop {\rm and \phantom{l} twisted}}
   \oh (\chi_k^W + \tilde{\chi}_k^W)^{*} \oh (\chi_k^W + \tilde{\chi}_k^W)
+  \oh (\chi_k^W - \tilde{\chi}_k^W)^{*} \oh (\chi_k^W - \tilde{\chi}_k^W).
\cr
}    \eqno({\rm 3.1.6})$$
This implies that the characters of the orbifold $\w$-algebra are given
by $\oh (\chi_k^W + \tilde{\chi}_k^W)$ and
$\oh (\chi_k^W - \tilde{\chi}_k^W)$. We conclude that the $h$-values
for the HWRs of the orbifold are those of the original $\w$-algebra
in both sectors in addition to some which differ by (half-) integers.
\sn
It is a special feature of the $\w$-algebras discussed in this section
that orbifolding yields just the Virasoro algebra. In general, the
orbifold will also be non-linear and will have more generators than
the original algebra. The identification of partition functions (3.1.6)
will be valid for all $\w$-algebras with a $\Zed_2$-automorphism.
\bn
\leftline{\bf 3.2.\ Parabolic $\ww$-algebras}
\mn
There is another series of $\w(2, \delta)$-algebras with vanishing
self coupling constant that leads to rational models $\q{\wirrep}$.
Here, the relation $c=1 - 8 \delta$ holds $\q{\blm}$ and the effective
central charge satisfies $\tilde{c} = 1$ $\q{\wirrep}$ (therefore these models
are called `parabolic'). The bosonic members of this series that we have
studied explicitly are $\w(2,3)$ at $c=-23$ and $\w(2,6)$ at $c=-47$.
Define $m$ by $c=1-12 m$.
Then our result is that the relevant HWRs can be parametrized by
$$\eqalignno{
h_{c;{n \over 2 m},{n \over 2 m} } &= {n^2 \over 8 m}
   - {m \over 2}   &({\rm 3.2.1a}) \cr
h_{c;{n \over 2 m + 4},-{n \over 2 m + 4} }
   &= {n^2 \over 8 m +16} - {m \over 2}   &({\rm 3.2.1b})\cr
}$$
with $n \in \Zed_{+}$, $n \leq m$ and $n = 2m$. For the bosonic algebras
even $n$ yields the untwisted sector of the algebra while odd $n$ leads to the
twisted sector. This parallels the structure of the fermionic members
of this series ($\w(2,{9 \over 2})$ at $c=-35$ and $\w(2,{15 \over 2})$
at $c=-59$) where even and odd $n$ yield the Neveu-Schwarz- respectively
Ramond-sector. As soon as the twisted sector of bosonic $\w$-algebras is taken
into account all members of this
series are described by the same formula (3.2.1) and no different cases
(as in $\q{\wirrep}$) have to be considered.
\sn
These explicit results being present, the characters of the twisted
sector of these algebras also have been realized in terms of
Jacobi-Riemann-Theta functions and general arguments
have been given that (3.2.1) actually describes a complete series of
$\w(2, \delta)$-algebras $\q{\mfl}$.
\bn
\leftline{\bf 3.3.\ $\ww$-algebras related to non-minimal (1,\kk)-models}
\mn
In this section we want to discuss the last series of bosonic $\w$-algebras
that have vanishing self coupling constant. Here we can evaluate $c=c_{1,k}$
and $\delta = h_{1,k;1,3}$ from (3.1.1) (we use the parametrization of the
central charge as a name for this series). It was shown by H.G.\ Kausch
$\q{\kausch}$ that the series indeed exists and that it can be realized
in terms of free fields. Explicit examples for this series are $\w(2,3)$
at $c=-2$, $\w(2,5)$ at $c=-7$
and $\w(2,7)$ with $c=-{25 \over 2}$.
For the untwisted sector of these algebras two one-parameter families of
representations were obtained in $\q{\wirrep}$. This was a first hint
that these models are not rational. Since we know the vacuum
character for these algebras $\q{\nahmpriv}$:
$$
\chi_0^W = {1 \over \eta(q)}
\sum_{n \in \zeds} {\rm sign}(n) q^{{(2 k n + k -1)^2 \over 4 k}}
\eqno({\rm 3.3.1})$$
we could in principle calculate the orbit of $\chi_0^W$ under the modular
group and check if it is finite. However,
if one wants to rewrite (3.3.1) in terms of quadratic forms one is lead
almost immediately to infinite sums and it is not very plausible that this
orbit is finite. This provides us with a second argument that these models
are not rational.
\medskip
The fact that we can still use (3.1.1) for the parametrization of $c$ and
$\delta$ might seem strange for the conformal grid degenerates to a void set
at $p=1$. However, it does make sense to restrict to $r=1$ in the approach
of Felder $\q{\felder}$. In $\q{\wirrep}$ it was pointed out
that the representations of the untwisted sector of these algebras satisfy
a relation of the form
$$w^2 = \alpha_k \prod_{1 \le r \le 2 k -1 \atop r \in \zeds}
                 {\big (} h - h_{1,k;r,1}  {\big )} \eqno({\rm 3.3.2})$$
where the explicit values of $\alpha_k$ have been given in $\q{\wirrep}$.
However, this can be rewritten as
$$w^2 = \alpha_k \prod_{3 - 2 k \le n \le 2 k + 1 \atop n \in 2 \zeds}
                 {\big (} h - h_{1,k;1,{n \over 2 k}} {\big )}
.  \eqno({\rm 3.3.3})$$
If we now study the HWRs of the twisted sector of these algebras explicitly
we obtain from every condition in both approaches {\it finitely} many HWRs.
The results can be summarized in the following parametrization of the
$L_0$-eigenvalues:
$$h \in \{ h_{1,k;1,{n \over 2 k}} \mid
n \in 2 \Zed + 1, 3 - 2 k \le n \le 2 k + 1 \}. \eqno({\rm 3.3.4})$$
As these models are not rational, it is surprising that their twisted sector
is finite.
\sn
Obviously, (3.3.3) and (3.3.4) should generalize to all members of this
series.
\medskip
Recently, rational models have been discovered for $c_{1,k}$ $\q{\flohkau}$.
The results of $\q{\flohkau}$ confirm our observation that there are no
rational models at $c_{1,k}$ with a $\w(2, \delta)$ symmetry algebra because
the model becomes rational only if two currents are included in the
symmetry algebra. The algebras we discussed here may be considered as
subalgebras of the algebra containing currents. In this language, the
energy-momentum tensor $L$ and the additional simple field $W$
are presumably composite. A further interesting
property of the rational models at $c_{1,k}$ is that they are effective
$c=1$ theories $\q{\flohkau}$.
\bn
\leftline{\bf 4.\ Generalizations to Casimir algebras}
\mn
So far we have been discussing bosonic $\w(2, \delta)$-algebras.
The next step is to consider other, more general $\w$-algebras.
Most attention is attracted by
so-called `Casimir' algebras ${\cal WL}_n$ where
the dimension of the simple fields equals the order of the Casimir
invariants in a simple Lie algebra ${\cal L}_n$ $\q{\bbss} \q{\bai}$.
One common approach to their study is
Toda field theory $\q{\baf} \q{\bal} \q{\blg}$.
Their unitary minimal series can also be studied via GKO-constructions
$\q{\goddard}$. Q.\ Ho-Kim and H.B.\ Zheng have noticed that in this
approach outer automorphisms of the Lie algebra give rise to automorphisms of
the $\w$-algebra and argued that there no further ones $\q{\hokimA} \q{\hokimB}
\q{\hokimC}$. Owing to their work twists of the unitary minimal series of
Casimir algebras are well understood. Still, we would like to comment on
Casimir algebras from the point of view of extended conformal algebras,
especially in their non-unitary regime.
\mn
We have already stated that the Casimir algebras $\w(2,3) \cong {\cal WA}_2$
with generic $c$
and $\w(2,6) \cong {\cal WG}_2$ at $c=-{516 \over 13}$ and $c=-47$
have exactly one outer automorphism while the algebras
$\w(2,4) \cong {\cal WB}_2 \cong {\cal WC}_2$ as well as
$\w(2,6) \cong {\cal WG}_2$ with generic $c$ have no outer automorphism.
The next simple example is $\w(2,3,4) \cong {\cal WA}_3$. This algebra has
been explicitly constructed by R.\ Blumenhagen et al.\ $\q{\blm}$. The
explicit structure of the algebra (especially the vanishing of some coupling
constants) shows that this $\w$-algebra possesses exactly one outer
automorphism which is given by $V \mapsto  -V$ where $V$ is the simple field
of dimension 3. The algebra contains one null field of dimension 6 and one
of dimension 7 at $c=1$ and $c=-{116 \over 3}$. This enables us to study
the representation theory at these values of the central charge explicitly.
For both values of the central charge there are only finitely many
values of $h$ (with specific eigenvalues of the zero modes of the additional
bosonic fields which we omit here) for which these two null fields do
indeed vanish. They are listed in the following table:
\mn
\centerline{
\vrule \hskip 1pt
\vbox{ \offinterlineskip
\def\tablespace{height2pt&\omit&&\omit&&\omit&&\omit&&\omit&&\omit&&\omit&\cr}
\def\tablerule{ \tablespace
                \noalign{\hrule}
                \tablespace        }
\hrule
\halign{&\vrule#&
  \strut\hskip 4pt\hfil#\hfil\hskip 4pt\cr
height4pt& \multispan{13} & \cr
& \multispan{13} \hfil $\w(2,3,4) \cong {\cal WA}_3$ \hfil &\cr
height4pt& \multispan{13} & \cr
\noalign{\hrule}
height3pt & \multispan7 && \multispan5 &\cr
& \multispan7 \hfil $c=c_{4,9}^{\A_3}=-{116 \over 3}$ \hfil
     && \multispan5 \hfil $c=c_{5,6}^{\A_3}=1$ \hfil       &\cr
height3pt & \multispan7 && \multispan5 &\cr
\noalign{\hrule}
height2pt & \multispan3 && \multispan3 && \multispan3 && \omit &\cr
& \multispan3 \hfil {\it untwisted} \hfil && \multispan3 \hfil {\it twisted}
         \hfil
   && \multispan3 \hfil {\it untwisted} \hfil && {\it twisted}  &\cr
height2pt & \multispan3 && \multispan3 && \multispan3 && \omit &\cr
\noalign{\hrule}
\tablespace
&  $0$             &&  $-{13 \over 9}$ && $-{5 \over 4}$   && $-{59 \over 36}$
    && $0$            && ${9 \over 16}$ && ${1 \over 16}$   &\cr \tablespace
&  $-{4 \over 3}$  &&  $-{14 \over 9}$ && $-{17 \over 12}$ && \omit
    && $1$            && ${1 \over 16}$ && ${3 \over 16}$   &\cr \tablespace
&  $-{5 \over 3}$  &&  $-{2 \over 3}$  && $-{19 \over 12}$ && \omit
    && ${1 \over 3}$  && \omit          && ${1 \over 48}$   &\cr \tablespace
&  $-{8 \over 9}$  &&  $-1$            && $-{29 \over 36}$ && \omit
    && ${1 \over 12}$ && \omit          && ${25 \over 48}$  &\cr \tablespace
&  $-{11 \over 9}$ &&  \omit           && $-{53 \over 36}$ && \omit
    && ${3 \over 4}$  && \omit          &&    \omit         &\cr \tablespace
}
\hrule}\hskip 1pt \vrule
}
\mn
We have not included the solution $h={17 \over 32}$ in the twisted sector
at $c=1$ into the above table because we believe it to be a remnant that
would vanish if further conditions were studied.
\mn
The next algebra in the ${\cal WA}_n$-series is
$\w(2,3,4,5) \cong {\cal WA}_4$. It has been shown in $\q{\hornfeck}$
that there are two solutions for a $\w$-algebra with additional simple fields
of dimension 3, 4 and 5 which we denote by $U$, $V$ and $S$.
For both solutions $C_{US}^V \neq 0$ holds. Thus, both solutions -- in
particular the one corresponding to ${\cal WA}_4$ -- have exactly one outer
automorphism which is given by $U \mapsto -U$ and $S \mapsto -S$.
\medskip
{}From $\q{\hokimA} \q{\hokimB} \q{\hokimC}$ we know that for the unitary
minimal series of Casimir algebras the number of automorphisms of the
$\w$-algebra coincides with that of the corresponding Lie algebra.
This is also clear in the Fateev-Lykyanov-construction $\q{\lykyanov}$
of these algebras.
If one demands the additional fields to be primary one has to add correction
terms to the pure Casimir invariant terms. It is highly non-trivial that these
respect the automorphisms of the underlying Lie algebra. However,
we explicitly observed that ${\cal WA}_2$, ${\cal WA}_3$ and ${\cal WA}_4$
have exactly one outer automorphism for generic value of the central charge
$c$ while for ${\cal WA}_1$, ${\cal WB}_2 \cong {\cal WC}_2$ and ${\cal WG}_2$
there are none.
For these cases the number of automorphisms of ${\cal WL}_n$ coincides
with those of ${\cal L}_n$ for
generic value of the central charge $c$. Consequently, we expect that
${\cal WL}_n$ generically has as many automorphisms as ${\cal L}_n$ has.
This observation is equivalent to a covariance property of a free field
construction which has up to now never been used although one automorphism
reduces the number of unknowns by one half in such a construction.
\mn
Thus, there will be no outer automorphisms
for ${\cal WA}_1$, ${\cal WB}_n$, ${\cal WC}_n$, ${\cal WE}_7$, ${\cal WE}_8$,
${\cal WF}_4$ and ${\cal WG}_2$. The algebras ${\cal WA}_n$ for $n > 1$,
${\cal WD}_n$ for $n > 4$ and ${\cal WE}_6$ should have exactly one
outer automorphism and correspondingly exactly one twisted sector in
addition to the untwisted one.
For these algebras our explicit data shows how the formulae for the
$h$-values in $\q{\hokimB} \q{\hokimC}$
might generalize:
$$\eqalign{
c_{p,q}^{\Lie_n} &= n - 12 \rho^2 {(p-q)^2 \over p q} \ , \cr
h_{p,q;\lambda,\mu}^{\Lie_n^{(2,3)}} &=
   {(p \lambda - q \mu)^2 \over 2 p q}+{c_{p,q}^{\Lie_n}-n \over 24}
               + \tilde{h} n_1 \ , \cr
}\eqno({\rm 4.1})$$
where
$$\rho = \sum_{i=1}^{n} \tilde{\Omega}_i \  , \qquad
\lambda = \sum_{i=1}^{n_0} r_i \Omega_i \  , \qquad
\mu = \sum_{i=1}^{n_0} s_i \Omega_i \  ,
\eqno({\rm 4.2})$$
and $n_0$, $n_1$ are the dimensions of the invariant subalgebra
$\hat{\Lie}_{n_0}$ respectively twisted subalgebra of $\Lie_n$;
$\tilde{\Omega}_i$ are the fundamental weights of $\Lie_n$;
$\Omega_i$ the fundamental weights of $\hat{\Lie}_{n_0}$; $\tilde{h}$
the conformal dimension of the twisted field and $r_i$, $s_i$ positive
integers subject to certain constraints.
For the case of $\A_n$ the invariant subalgebra is
$\AC_{\lbrack {n+1 \over 2} \rbrack}$. In the unitary minimal series of
${\cal WA}_n$, ${\cal WD}_n$ ($n > 4$) and ${\cal WE}_6$
one has $\tilde{h} = {1 \over 16}$. We observe that (4.1) indeed
reproduces our explicit data for $\w(2,3,4)$ if we use the weights
of $\AC_2$ (which is the invariant subalgebra) and
$\tilde{h} = {1 \over 16}$ not only in the unitary case but also in the
non-unitary case. It should be possible to prove (4.1) with (4.2)
rigorously applying quantized Drinfeld-Sokolov reduction to the
characters of $\Lie_n^{(2)}$ and $\D_4^{(3)}$, thus generalizing the work
of $\q{\frenkel}$ on $\Lie_n^{(1)}$.
\medskip
The exceptional cases ${\cal WE}_6$ and ${\cal WD}_4$ are particularly
interesting. Especially for ${\cal WD}_4 \cong \w(2,4,4,6)$
the group of outer automorphisms should be $\S_3$ for generic $c$.
This case
is up to now the only case where more than just one outer automorphism is
known. In $\q{\blm}$ and $\q{\kau}$ the algebra $\w(2,4,4)$ has been
shown to
be consistent for $c=1$ and $c=-{656 \over 11}$. It has then been conjectured
in $\q{\kausch}$ that the Casimir algebra of $\D_4$ should reduce to
$\w(2,4,4)$ for these two values of the central charge.
For $\w(2,4,4)$ one structure constant remains free such that one can define
an operation of $O(2)$ on the two additional fields under which
the structure constants transform covariant.
{}From the explicit form of the structure constants given in $\q{\kausch}$
one sees that the algebra is
invariant under the natural embedding of $\S_3$ into the group $O(2)$
operating on the fields if one chooses the self coupling
constants of the two additional simple fields to be equal.
In fact, this is also true for $\w(2,4,4,6)$ $\q{\hgkpriv}$.
\sn
Denote the primary fields of dimension 4 in $\w(2,4,4,6)$ by $V(z)$ and
$W(z)$. Then the $\S_3$-symmetry of this algebra translates into the
following type of boundary conditions
for the given choice of coupling constants:
$$\eqalign{
V(e^{2 \pi i} z) &= \cos(\alpha) V(z) - \sin(\alpha) W(z) \cr
W(e^{2 \pi i} z) &= \sin(\alpha) V(z) + \cos(\alpha) W(z) \cr
} \eqno({\rm 4.3})$$
or
$$\eqalign{
V(e^{2 \pi i} z) &= \cos(\alpha) W(z) - \sin(\alpha) V(z) \cr
W(e^{2 \pi i} z) &= \sin(\alpha) W(z) + \cos(\alpha) V(z) \cr
} \eqno({\rm 4.4})$$
with $\alpha \in \{ 0, {2 \over 3} \pi, {4 \over 3} \pi \}$.
The three different boundary conditions given by (4.3)
correspond to those elements of $\S_3$ which under the embedding
yield elements of $SO(2)$. The boundary conditions (4.4) correspond to
the three elements of $\S_3$ that are mapped to elements in $O(2)$ with
determinant $-1$.
\sn
The $h$-values in the unitary minimal series (4.1) of ${\cal WD}_4$
have been calucated in $\q{\hokimB}$ without having to consider the
boundary conditions of the additional simple fields which look quite strange
at first sight. Note that for
$\D_4$ the invariant subalgebra is the exceptional algebra $\G_2$
and here the dimension of the twisted field is
$\tilde{h} = {1 \over 18}$.
\sn
In $\q{\bouwschou}$ it has already been stated that the $\S_3$-symmetry
should lead to modes in ${\zed \over 3}$. We shall show now that this is
indeed correct, and we will discuss how the modes have to be chosen precisely.
Let us first focus on (4.3). Set $U^{(1)}(z) := V(z) + i W(z)$ and
$U^{(2)}(z) := V(z) - i W(z)$. Then (4.3) turns into
$U^{(1)}(e^{2 \pi i} z) = e^{i \alpha} U^{(1)}(z)$ and
$U^{(2)}(e^{2 \pi i} z) = e^{-i \alpha} U^{(2)}(z)$ which can be satisfied
by choosing modes in $\Zed + {\alpha \over 2 \pi}$ for $U^{(1)}$ and
those for $U^{(2)}$ in $\Zed - {\alpha \over 2 \pi}$.
Consider now (4.4). For this case set
$Y^{(1)}(z) := \cos(\alpha) V(z) + (\sin(\alpha)+1) W(z)$ and
$Y^{(2)}(z) := \cos(\alpha) V(z) + (\sin(\alpha)-1) W(z)$. Now (4.4)
turns into $Y^{(1)}(e^{2 \pi i} z) = Y^{(1)}(z)$ and
$Y^{(2)}(e^{2 \pi i} z) = -Y^{(2)}(z)$. This can be satisfied by
choosing the modes for $Y^{(1)}$ in $\Zed$ and those for $Y^{(2)}$
in $\Zed+{1 \over 2}$.
\medskip
However, this procedure
shows that outer automorphisms of a $\w$-algebra can lead to
even more complicated structures than discussed in this paper. Nevertheless,
for the case of ${\cal WA}_n$ the outer automorphism will operate
as reflection on the space of primary fields with odd dimension. Therefore,
in the twisted sector of ${\cal WA}_n$ all primary fields with odd dimension
will obtain half-integral modes and the representation theory will follow
the lines of this paper.
\mn
As we have seen for $\w(2,6) \cong {\cal WG}_2$ there may be specific values
of the central charge where some structure constans vanish and thus enlarge
the automorphism group. These phenomena are very interesting but in order to
discuss them generally one would have to know the structure constants of all
Casimir algebras which goes beyond current knowledge.
\medskip
It would be interesting to generalize the observations of this chapter
to the supersymmetric case. Of special interest is $osp(4 \mid 4)$
which is the supersymmetric analogon of $\D_4$.
The corresponding Casimir algebra is a
${\cal SW}({3 \over 2}, 2, 2, {7 \over 2})$. In $\q{\supwir}$ the algebra
${\cal SW}({3 \over 2}, 2, 2)$ has been shown to be consistent only for
$c={3 \over 2}$ and one may expect this algebra to coincide with the
Casimir algebra of $osp(4 \mid 4)$ for this specific value of the central
charge. As in the case of $\w(2,4,4)$ one structure constant remains free. For
${\cal SW}({3 \over 2}, 2, 2)$ it has been noticed in $\q{\supwir}$
that this algebra is invariant under the natural embedding of $\S_3$ into
the group $O(2)$ operating on the fields if one chooses the self coupling
constants of the two additional simple fields to be equal. Thus,
${\cal SW}({3 \over 2}, 2, 2, {7 \over 2})$ should also admit boundary
conditions similar to (4.3), (4.4).
\bn
\leftline{\bf 5.\ Applications in statistical mechanics}
\mn
As already pointed out in $\q{\hokimA}$ there is a close connection
of the partition function of $\w(2,3)$ at $c={4 \over 5}$
including the twisted sector and the three states Potts model. In fact,
the different boundary conditions of $\w(2,3)$ correspond to the different
boundary conditions of the Potts quantum spin chain at critical temperature.
Choosing the spin shift operator at the end of the chain to be equal to the
spin shift operator at the first site yields the field content of the
`untwisted' sector of $\w(2,3)$ (see e.g.\ $\q{\fateev} \q{\zamzam}$ and
$\q{\gehlen}$ for numerical verification). We recall the remarkable
fact $\q{\hokimA}$ that the
twisted sector of $\w(2,3)$ yields additional representations which can be
identified
with fields in the thermodynamic limit of the three states Potts quantum
spin chain if the spin shift operator at the end of the chain is chosen
to equal the adjoint of the one at the first site $\q{\cardy}$.
This has been verified by Cardy using the inversion identity method.
In $\q{\cardy}$ he called this type of boundary conditions also
for the statistical mechanics model `twisted'.
\mn
It would be interesting to know if this observation generalizes to
{\it all} $\Zed_n$. For twisted boundary conditions only partial
results are available in the literature (see e.g.\ $\q{\schuetz}$).
We shall therefore present an explicit verification
of this statement in the case of $\Zed_4$. We will
follow the approach of $\q{\gehlen}$ and study the spectrum
of the following hamiltonian numerically:
$$H^{(n)}_N = - {1 \over n}
                   \sum_{j=1}^N \sum_{k=1}^{n-1}
                       {1 \over \sin{\pi k \over n}} {\bigg (} \si_j^k
                  + \la \Ga_j^k \Ga_{j+1}^{n-k} {\bigg )},
             \eqno{(\rm 5.1)}$$
where $\si_j$ and $\Ga_j$ freely generate a finite dimensional
associative algebra by the following relations ($1 \le j,l \le N$):
$$\eqalign{
\si_j \Ga_l = \Ga_l \si_j \om^{\delta_{j,l}}_{}, & \qquad
\si_j^n = \Ga_j^n = \id, \cr
\si_j \si_l = \si_l \si_j, & \qquad
\Ga_j \Ga_l = \Ga_l \Ga_j, \cr
} \eqno{(\rm 5.2)}$$
with $\om = e^{2 \pi i \over n}_{}$. One can impose different types
of boundary conditions for $H^{(n)}_N$. We will follow Cardy $\q{\cardy}$
and call the boundary condition $\Ga_{N+1} = \Ga_1$ `periodic' and
$\Ga_{N+1} = \om^{-R} \Ga_1$ ($R \ne 0$) `cyclic'.
Our main interest is in `twisted'
boundary conditions which are given by $\Ga_{N+1} = \Ga_1^{+}$.
Finally, the spin quantum chain (5.1) also admits boundary conditions of
the form $\Ga_{N+1} = \om^{-R} \Ga_1^{+}$ ($R \ne 0$).
\sn
It is convenient to use an irreducible representation of the algebra (5.2)
in $\otimes^N \Complex^n$ where $\si_j$ and $\Ga_j$ are represented
in terms of diagonal respective shift matrices.
\sn
Let $E_{N,i}$ be the eigenvalues of $H_N^{(n)}$ with periodic boundary
conditions in ascending order and $\tilde{E}_{N,i}$ those with twisted
boundary conditions. Then the relevant scaling functions are given by
$\q{\cardyA} \q{\gehlenA}$:
$$\eqalign{
\xi_{N,i} &:= {N \over 2 \pi} (\tilde{E}_{N,i} - E_{N,0}) \cr
\xi_i &:= \lim_{N \to \infty} \xi_{N,i}. \cr
}  \eqno({\rm 5.3})$$
In the case of periodic and cyclic boundary conditions, the
eigenvalue $\om^Q$ ($Q=0,\ldots, n-1$) of the charge operator
$\hat{Q} = \prod_{j=1}^N \si_j$ and momentum are good
quantum numbers. In the case of twisted boundary conditions
neither charge nor momentum are conserved any more and one
does not have any obvious conserved quantities.
At least for even $n$ the charge $Q \mod 2$ is conserved.
\mn
Assume that (5.1) exhibits conformal invariance at $\la=1$ and
denote the dimensions of the fields in the left chiral
part by $h$ and of those in the right chiral part by
$\bar{h}$. For periodic and twisted boundary conditions
the field theory is diagonal, i.e.\
the fields $\phi(z, \bar{z})$ with dimension $h+\bar{h}$
satisfy $h=\bar{h}$ and thus have vanishing spin $h-\bar{h}$.
Therefore, the modes of the fields $\phi(z, \bar{z})$
yield levels in the spectrum with $\xi = h+\bar{h}+r$ where
$r \in \Zed_{+}$ for periodic boundary conditions and
$r \in {\zed_{+} \over 2}$ for twisted boundary conditions.
\mn
In order to test this method we shall first study the well
known three states Potts model.
For $\Zed_3$ we have studied $3$ to $9$ sites. Thus, it was
necessary to partially diagonalize matrices of dimension
$3^9 = 19683$. The limits $N \to \infty$ of the lowest gaps
$\xi_i$ are given in the following table:
\mn
\centerline{
\vrule \hskip 1pt
\vbox{ \offinterlineskip
\def\tablespace{ height2pt&\omit&&\omit&&\omit&\cr }
\def\tablerule{ \tablespace
                \noalign{\hrule}
                \tablespace        }
\hrule
\halign{&\vrule#&
  \strut\hskip 4pt\hfil#\hfil\hskip 4pt\cr
height4pt& \multispan{5} & \cr
& \multispan{5} \hfil $\Zed_3$ \hfil &\cr
height4pt& \multispan{5} & \cr
\noalign{\hrule}
\tablespace
& $i$   &&  $\xi_i$      &&  $h+\bar{h}+r$    &\cr
\tablespace
\noalign{\hrule}
\tablespace
& $0$   && $0.050000(2)$ && ${1\over 40}+{1\over 40}$  &\cr\tablespace
& $1$   && $0.250005(5)$ && ${1\over 8}+{1\over 8}$    &\cr\tablespace
& $2$   && $0.5500(5)$   && ${1\over 40}+{1\over 40}+{1\over 2}$&\cr\tablespace
& $3$   && $1.05(4)$     && ${1\over 40}+{1\over 40}+1$  &\cr\tablespace
}
\hrule}\hskip 1pt \vrule
}
\mn
The numbers in brackets indicate the estimated error in the
last given digit. For details on the extrapolation procedures
and error estimation see e.g.\ $\q{\henkel}$.
We do not give more than four levels because
the errors of the next levels make an accurate identification
impossible.
Note that we can nicely identify the dimensions
${1 \over 40}$ and ${1 \over 8}$ of the chiral fields -- as expected.
\medskip
Let us now turn to $\Zed_4$. The spectrum
of the $\Zed_4$ chain has already been derived in $\q{\baakeB}$
also for twisted boundary condition. This was done applying
numerical and Kac-Moody algebra techniques. Nonetheless, we will
present results of a direct calculation here because we would like
to demonstrate the correspondence between boundary conditions
in statistical mechanics and conformal field theory.
Note that the $\Zed_4$-version
of (5.1) is a special case of the Ashkin-Teller quantum chain which was
introduced in $\q{\kohmoto}$ setting the parameter $h={1 \over 3}$
(in the notations of $\q{\baakeA}$).
\sn
For $\Zed_4$ we have at least a splitting of the spectrum into two
sectors of $Q \mod 2$. We have studied $4$ to $8$ sites, implying
the partial diagonalization of matrices in dimensions up to
${4^8 \over 2} = 32768$.
\mn
\centerline{
\vrule \hskip 1pt
\vbox{ \offinterlineskip
\def\tablespace{
height2pt&\omit&&\omit&&\omit&\hskip 1pt \vrule&\omit&&\omit&&\omit&\cr }
\def\tablerule{ \tablespace
                \noalign{\hrule}
                \tablespace        }
\hrule
\halign{&\vrule#&
  \strut\hskip 4pt\hfil#\hfil\hskip 4pt\cr
height4pt& \multispan{11} & \cr
& \multispan{11} \hfil $\Zed_4$ \hfil &\cr
height4pt& \multispan{11} & \cr
\noalign{\hrule}
height3pt & \multispan5 &\hskip 1pt \vrule& \multispan5 &\cr
& \multispan5 \hfil $Q \mod 2 = 0$ \hfil  &\hskip 1pt \vrule&
          \multispan5 \hfil $Q \mod 2 = 1$ \hfil       &\cr
height3pt & \multispan5 &\hskip 1pt \vrule& \multispan5 &\cr
\noalign{\hrule}
\tablespace
& $i$   &&  $\xi_i$     &&  $h+\bar{h}+r$  &\hskip 1pt \vrule&
    $i$   &&  $\xi_i$    &&  $h+\bar{h}+r$  &\cr
\tablespace
\noalign{\hrule}
\tablespace
& $0$   && $0.04167(2)$ && ${1\over 48}+{1\over 48}$ &\hskip 1pt \vrule&
     $0$  &&  $0.1254(1)$ && ${1\over 16}+{1\over 16}$ &\cr\tablespace
& $1$   && $0.375(2)$   && ${3\over 16}+{3\over 16}$ &\hskip 1pt \vrule&
     $1$  &&  $0.6231(3)$ && ${1\over 16}+{1\over 16}+{1\over 2}$
              &\cr\tablespace
& $2$   && $1.040(2)$   && ${25\over 48}+{25\over 48}$ &\hskip 1pt \vrule&
     $2$  &&  $0.623(7)$  && ${1\over 16}+{1\over 16}+{1\over 2}$
              &\cr\tablespace
& \omit && \omit        && \omit &\hskip 1pt \vrule&
     $3$  &&  $1.12(1)$   && ${1\over 16}+{1\over 16}+1$ &\cr\tablespace
}
\hrule}\hskip 1pt \vrule
}
\mn
The dimensions ${1 \over 48}$, ${1 \over 16}$, ${3 \over 16}$ and
${25 \over 48}$ of the chiral field theory can be nicely seen in
these explicit results.
\medskip
It is well known that the $\Zed_n$-models (5.1) with periodic boundary
conditions at their second
order phase transition $\la = 1$ exhibit a ${\cal WA}_{n-1} \cong
\w(2,\ldots,n)$-symmetry $\q{\lykyanov} \q{\alcaraz}$.
The field content of this model is given by
the first unitary representation of ${\cal WA}_{n-1}$, i.e.\ by
(4.1) with $\hat{\Lie}_{n_0-1} = \Lie_{n-1} = \A_{n-1}$ and $p=n+1$,
$q=n+2$. In particular, the central charge for a $\Zed_n$-model (5.1)
equals $c = {2 (n-1) \over n+2}$.
In this section we have explicitly verified for
$n=3$ and $4$ that the representations of the twisted sector
of ${\cal WA}_{n-1}$ correspond to the spectrum of the
$\Zed_{n}$-model with twisted boundary conditions. In fact, this is
also true for the $\Zed_5$-version of (5.1) $\q{\gehlenunpub}$.
These explicit results are in agreement with the statement that
the field content of the spin quantum chain (5.1)
at $\la=1$ with twisted boundary conditions $\Ga_{N+1} = \Ga_1^{+}$
can be described by a representation of a twisted ${\cal WA}_{n-1}$
for all $n$. Thus, it is possible to calculate
the spectrum of the twisted $\Zed_n$-quantum chain by
(4.1) using $\Lie_{n-1} = \A_{n-1}$,
$\hat{\Lie}_{n_0} = \AC_{\lbrack {n \over 2} \rbrack}$
and $p=n+1$, $q=n+2$.
\mn
For cyclic boundary conditions ($\Ga_{N+1} = \om^{-R} \Ga_1$,
$0 < R < n$) the diagonal symmetry of the statistical mechanics
model is known to be broken such that the spin $h-\bar{h}$
takes on rational values. The dimensions of the chiral fields,
however, are unaffected by this change of boundary conditions.
This has been verified in $\q{\cardy}$ and $\q{\gehlen}$ for
the case $n=3$, in $\q{\baakeB}$ for $n=4$ and more abstractly
for general $n$ in $\q{\gepner}$.
In $\q{\cardy}$ a similar result has been obtained
for $\Ga_{N+1} = \om^{-R} \Ga_1^{+}$, $0 < R < 3$, $n=3$
and the only effect of a factor $\om^{-R}$
for all $n$ should be to combine the left- and right-chiral parts
in a non-diagonal way.
\bigskip
Finally, we should stress that also the $(1,k)$-models may have
important applications in statistical mechanics. In particular,
$c_{1,2} = -2$ turns up in the context of two-dimensional polymers
$\q{\saleur} \q{\duplantier}$. In the dense phase of two-dimensional
polymers, the surface exponents
correspond to the conformal dimensions of the fields in the complete
chiral algebra whereas the representations of this algebra can
be identified with the bulk exponents $\q{\saleur}$. In particular,
the twisted representations we discussed in this paper turn up
naturally when the bulk exponents are evaluated $\q{\duplantier}$.
\bn
\leftline{\bf 6.\ Conclusion}
\mn
We have shown that for all bosonic $\w(2,\delta)$-algebras with vanishing
self coupling constant one can impose anti-periodic boundary conditions on
the additional field. From the point of view of representations this
leads to an additional twisted sector of these algebras. The two sectors
of these algebras parallel much the Neveu-Schwarz- and Ramond-sector of
fermionic $\w(2,\delta)$-algebras (which always have vanishing self coupling
constant). Specifically, for the bosonic $\w$-algebras coming from the
ADE-classification as well as for those with $c=1 - 8 \delta$ the
representation theory of both sectors is much alike that of the fermionic
theory.
For the algebras with $c=1-8 \delta$ this has recently been well understood
in the work of M.\ Flohr $\q{\mfl}$.
\sn
Unfortunately, we have not been able to obtain any new modular
invariant partition function although the rational models enlarge when
adding the twisted sector.
\sn
The third and last series of bosonic $\w(2, \delta)$-algebras with a twisted
sector is not rational. Here $c=c_{1,k}$ and $h=h_{1,k;1,3}$ holds
and H.G.\ Kausch has shown that a free field construction for these algebras
is possible $\q{\kausch}$.
We have shown that the twisted sector of these algebras is finite.
Recently, rational models have been discovered at $c_{1,k}$ where the
symmetry algebras contains currents $\q{\flohkau}$. The $\w$-algebras we
considered here can be interpreted as subalgebras of these larger
symmetry algebras. In particular, the representations of the twisted
$\w(2, \delta)$-algebras in this series should lead to chiral fields
in the rational models.
\sn
At least the representations of the first member of this series
-- $\w(2,3)$ at $c=-2$ -- seem to be closely related to two-dimensional
polymer physics $\q{\saleur} \q{\duplantier}$, and in particular
the representations of the twisted sector turn up naturally in
this context.
\mn
We have argued und verified in the first cases that Casimir algebras
${\cal WL}_n$ have generically exactly as many outer automorphisms as
the underlying Lie algebra $\Lie_n$. In particular, all
algebras of type ${\cal WA}_n$ ($n > 1$) possess
exactly one twisted sector. For the unitary minimal series this observation
is in agreement with $\q{\hokimB} \q{\hokimC}$.
On the basis of explicit results we were able to
conjecture a general formula (4.1) for the $h$-values in their complete
minimal series. We have pointed out that the algebra ${\cal WD}_4 \cong
\w(2,4,4,6)$ should be particularly interesting because the group of outer
automorphisms is $\S_3$, leading to the possibility to choose modes in
${\zed \over 3}$.
\mn
Finally, we have shown that the representations of the twisted
sector of these algebras have applications in statistical mechanics.
Generally, the first unitary minimal model of ${\cal WA}_n$ is
closely related related to a second order phase transition in a
$\Zed_{n+1}$-model. The twist of the symmetry algebra corresponds to
twisted boundary conditions in the $\Zed_{n+1}$-model. This has been
explicitly verified for the three states Potts model which is described
by $\w(2,3)$ at $c={4 \over 5}$ and the Ashkin-Teller model at a special
parameter value which corresponds to $\w(2,3,4)$ at $c=1$.
\bn
\leftline{\bf Acknowledgments}
\mn
I am indebted to M.\ Terhoeven for substantial help and careful reading
of the manuscript. It is a pleasure to thank everybody working
at the `Physikalisches Institut' Bonn, Germany about conformal field
theory for the inspiring atmosphere and many useful discussions.
In particular, I would like to thank W.\ Eholzer, M.\ Flohr and R.\ H\"ubel
for continuous collaboration in the field of $\w$-algebras and W.\ Nahm
and G.v.\ Gehlen for constant support.
\vfill
\eject
\leftline{\bf References}
\mn
\settabs\+&\phantom{---------}&\phantom{
------------------------------------------------------------------------------}
& \cr
\+ &$\q{\bpz}$ & A.A.\ Belavin, A.M.\ Polyakov, A.B.\ Zamolodchikov & \cr
\+ &           & {\it Infinite Conformal Symmetry in Two-Dimensional Quantum
                   Field Theory}  & \cr
\+ &           & Nucl.\ Phys.\ {\bf B241} (1984) p.\ 333  & \cr
\+ & $\q{\bouwschou}$
               & P.\ Bouwknegt, K. Schoutens, {\it $\w$-Symmetry in Conformal
                   Field Theory}  & \cr
\+ &           & preprint CERN-TH.6583/92 (1992),
                   to be published in Physics Reports & \cr
\+ &$\q{\nahm}$
               & W.\ Nahm, {\it Chiral Algebras of Two-Dimensional Chiral
                   Field Theories and Their} & \cr
\+ &           & {\it Normal Ordered Products},
                   Proceedings Trieste Conference on & \cr
\+ &           & Recent Developments in Conformal Field Theories,
                   ICTP, Trieste (1989) p.\ 81 & \cr
\+ &$\q{\nam}$ & W.\ Nahm, {\it Conformal Quantum Field Theories in Two
                   Dimensions} & \cr
\+ &           & World Scientific, to be published & \cr
\+ &$\q{\kau}$ & H.G.\ Kausch, G.M.T.\ Watts, {\it A Study of $\w$-Algebras
                   Using Jacobi Identities}& \cr
\+ &           & Nucl.\ Phys.\ {\bf B354} (1991) p.\ 740 & \cr
\+ &$\q{\blm}$ & R.\ Blumenhagen, M.\ Flohr, A.\ Kliem,
                    W.\ Nahm, A.\ Recknagel, R.\ Varnhagen & \cr
\+ &           & {\it $\w$-Algebras with Two and Three Generators},
                    Nucl.\ Phys.\ {\bf B361} (1991) p.\ 255 & \cr
\+ &$\q{\rva}$ & R.\ Varnhagen, {\it Characters and Representations of
                   New Fermionic $\w$-Algebras} & \cr
\+ &           & Phys.\ Lett.\ {\bf B275} (1992) p.\ 87 & \cr
\+ &$\q{\wirrep}$
               & W.\ Eholzer, M.\ Flohr, A.\ Honecker,
                   R.\ H{\"u}bel, W.\ Nahm, R.\ Varnhagen  & \cr
\+ &           & {\it Representations of $\w$-Algebras with Two Generators
                   and New Rational Models } & \cr
\+ &           & Nucl.\ Phys.\ {\bf B383} (1992) p.\ 249 & \cr
\+ &$\q{\mfl}$ & M.\ Flohr, {\it $\w$-Algebras,
                   New Rational Models and Completeness of the $c=1$} & \cr
\+ &           & {\it Classification}, preprint BONN-HE-92-08 (1992) & \cr
\+ &$\q{\wowoA}$
               & W.\ Eholzer, {\it Fusion Algebras Induced by Representations
                   of the Modular Group} & \cr
\+ &           & preprint BONN-HE-92-30 (1992), hep-th/9210040 & \cr
\+ &           & to be published in Int.\ Jour.\ of Mod.\ Phys.\ {\bf A} & \cr
\+ &$\q{\wowoB}$
               & W.\ Eholzer, N.\ Skoruppa, {\it Exceptional
                  $\w$-Algebra Characters and Theta-Series} & \cr
\+ &           & {\it of Quaternion Algebras}, in preparation & \cr
\+ & $\q{\hokimA}$
               & Q.\ Ho-Kim, H.B.\ Zheng, {\it Twisted Conformal
                   Field Theories with $\Zed_3$ Invariance} & \cr
\+ &           & Phys.\ Lett.\ {\bf B212} (1988) p.\ 71 & \cr
\+ &$\q{\zam}$ & A.B.\ Zamolodchikov & \cr
\+ &           & {\it Infinite Additional Symmetries in Two-Dimensional
                   Conformal Quantum Field}  & \cr
\+ &           & {\it Theory}, Theor.\ Math.\ Phys.\ 65 (1986) p.\ 1205  & \cr
\+ &$\q{\kausch}$
               & H.G.\ Kausch & \cr
\+ &           & {\it Extended Conformal Algebras Generated by a Multiplet
                   of Primary Fields} & \cr
\+ &           & Phys.\ Lett.\ {\bf B259} (1991) p.\ 448 & \cr
\+ &$\q{\dijkgraaf}$
               & R.\ Dijkgraaf, C.\ Vafa, E.\ Verlinde, H.\ Verlinde & \cr
\+ &           & {\it The Operator Algebra of Orbifold Models} & \cr
\+ &           & Comm.\ Math.\ Phys.\ 123 (1989) p.\ 485 & \cr
\+ & $\q{\romans}$
               & L.J.\ Romans, {\it Realisations of Classical
                  and Quantum $\w_3$ Symmetry} & \cr
\+ &           & Nucl.\ Phys.\ {\bf B352} (1991) p.\ 829 & \cr
\+ &$\q{\bbss}$
               & F.A.\ Bais, P.\ Bouwknegt, M.\ Surridge, K.\ Schoutens & \cr
\+ &           & {\it Extensions of the Virasoro Algebra Constructed
                  from Kac-Moody Algebras Using} & \cr
\+ &           & {\it Higher Order Casimir Invariants},
                  Nucl.\ Phys.\ {\bf B304} (1988) p.\ 348 & \cr
\+ &$\q{\bai}$ & F.A.\ Bais, P.\ Bouwknegt, M.\ Surridge, K.\ Schoutens & \cr
\+ &           & {\it Coset Construction for Extended Virasoro Algebras} & \cr
\+ &           & Nucl.\ Phys.\ {\bf B304} (1988) p.\ 371 & \cr
\+ & $\q{\deckmyn}$
               & A.\ Deckmyn, S.\ Schrans,
                  {\it $\w_3$ Constructions on Affine Lie Algebras} & \cr
\+ &           & Phys.\ Lett.\ {\bf B274} (1992) p.\ 186 & \cr
\+ &$\q{\wirsuperwrep}$
               & W.\ Eholzer, A.\ Honecker, R.\ H{\"u}bel & \cr
\+ &           & {\it Representations of $N=1$ Extended
                   Superconformal Algebras with Two Generators} & \cr
\+ &           & Preprint BONN-HE-92-28 (1992), hep-th/9209030 & \cr
\+ &$\q{\bil}$ & A.\ Bilal, {\it Introduction to $\w$-Algebras},
                   preprint CERN-TH.6083/91 (1991) & \cr
\+ &$\q{\watts}$
               & G.M.T.\ Watts, {\it Determinant Formulae for
                   Extended Algebras in Two-Dimensional}& \cr
\+ &           & {\it Conformal Field Theory},
                   Nucl.\ Phys.\ {\bf B326} (1989) p.\ 648 & \cr
\+ & $\q{\fateev}$
               & V.A.\ Fateev, A.B.\ Zamolodchikov & \cr
\+ &           & {\it Conformal Quantum Field Theory Models in Two
                    Dimensions Having} & \cr
\+ &           & {\it $\Zed_3$ Symmetry},
                    Nucl.\ Phys.\ {\bf B280} (1987) p.\ 644 & \cr
\+ & $\q{\frenkel}$
               & E.\ Frenkel, V.\ Kac, M.\ Wakimoto & \cr
\+ &           & {\it Characters and Fusion Rules for $\w$-Algebras
                    via Quantized Drinfeld-Sokolov Reductions} & \cr
\+ &           & Comm.\ Math.\ Phys.\ 147 (1992) p.\ 295 & \cr
\+ & $\q{\hokimB}$
               & Q.\ Ho-Kim, H.B.\ Zheng, {\it Twisted Structures of
                   Extended Virasoro Algebras} & \cr
\+ &           & Phys.\ Lett.\ {\bf B223} (1989) p.\ 57 & \cr
\+ & $\q{\hokimC}$
               & Q.\ Ho-Kim, H.B.\ Zheng, {\it Twisted Characters
                    and Partition Functions in} & \cr
\+ &           & {\it Extended Virasoro Algebras},
                    Mod.\ Phys.\ Lett.\ {\bf A5} (1990) p.\ 1181 & \cr
\+ & $\q{\commute} $
               & A.\ Honecker, {\it A Note on the Algebraic Evaluation
                    of Correlators in Local Chiral} & \cr
\+ &           & {\it Conformal Field Theory},
                    preprint BONN-HE-92-25 (1992), hep-th/9209029 & \cr
\+ &$\q{\diplom}$
               & A.\ Honecker, {\it Darstellungstheorie von $\w$-Algebren und
                  Rationale Modelle in der} & \cr
\+ &           & {\it Konformen Feldtheorie},
                  Diplomarbeit BONN-IR-92-09 (1992) &\cr
\+ &$\q{\cap}$ & A.\ Cappelli, C.\ Itzykson, J.B.\ Zuber & \cr
\+ &           & {\it The A-D-E Classification of Minimal and $A_1^{(1)}$
                  Conformal Invariant Theories} & \cr
\+ &           & Comm.\ Math.\ Phys.\ 113 (1987) p.\ 1 & \cr
\+ &$\q{\ver}$ & E.\ Verlinde & \cr
\+ &           & {\it Fusion Rules and Modular Transformations in
                   2D Conformal Field Theory} & \cr
\+ &           & Nucl.\ Phys.\ {\bf B300} (1988) p.\ 360 & \cr
\+ & $\q{\nahmpriv}$
               & W.\ Nahm, private communication & \cr
\+ & $\q{\felder}$
               & G.\ Felder, {\it BRST Approach to Minimal Models} & \cr
\+ &           & Nucl.\ Phys.\ {\bf B317} (1989) p.\ 215,
                   erratum Nucl.\ Phys.\ {\bf B324} (1989) p.\ 548 & \cr
\+ & $\q{\flohkau}$
               & M.\ Flohr, H.G.\ Kausch,
                  {\it A New Series of Rational Conformal Field Theories} & \cr
\+ &           & in preparation & \cr
\+ &$\q{\baf}$ & J.\ Balog, L.\ Feh\'er, L.\ O'Raifeartaigh,
                  P.\ Forg\'acs, A.\ Wipf & \cr
\+ &           & {\it Toda Theory and $\w$-Algebra from a Gauged WZNW
                  Point of View} & \cr
\+ &           & Ann.\ Phys.\ 203 (1990) p.\ 76 & \cr
\+ &$\q{\bal}$ & J.\ Balog, L.\ Feh\'er, P.\ Forg\'acs,
                  L.\ O'Raifeartaigh, A.\ Wipf & \cr
\+ &           & {\it Kac-Moody Realization of $\w$-Algebras} & \cr
\+ &           & Phys.\ Lett.\ {\bf B244} (1990) p.\ 435 & \cr
\+ &$\q{\blg}$ & A.\ Bilal, J.L.\ Gervais & \cr
\+ &           & {\it Systematic Construction of Conformal Theories with
                   Higher-Spin Virasoro} & \cr
\+ &           & {\it Symmetries}, Nucl.\ Phys.\ {\bf B318} (1989) p.\ 579 &\cr
\+ & $\q{\goddard}$
               & P.\ Goddard, A.\ Kent, D.\ Olive & \cr
\+ &           & {\it Unitary Representations of the Virasoro
                   and Super-Virasoro Algebras}  & \cr
\+ &           & Comm.\ Math.\ Phys.\ 103 (1986) p.\ 105 & \cr
\+ & $\q{\hornfeck}$
               & K.\ Hornfeck, {\it $\w$-Algebras with Set of
                   Primary Fields of Dimension}  & \cr
\+ &           & {\it (3,4,5) and (3,4,5,6)}, in preparation  & \cr
\+ & $\q{\lykyanov}$
               & V.A.\ Fateev, S.L.\ Lykyanov, {\it The Models of
                   Two-Dimensional Conformal} & \cr
\+ &           & {\it Quantum Field Theory with $\Zed_n$ Symmetry} & \cr
\+ &           & Int.\ Jour.\ of Mod.\ Phys.\ {\bf A3} (1988) p.\ 507 & \cr
\+ & $\q{\hgkpriv}$
               & H.G.\ Kausch, private communication & \cr
\+ &$\q{\supwir}$
               & R.\ Blumenhagen, W.\ Eholzer, A.\ Honecker, R.\ H{\"u}bel &\cr
\+ &           & {\it New N=1 Extended Superconformal Algebras
                  with Two and Three Generators } & \cr
\+ &           & preprint BONN-HE-92-02 (1992),
                  to be published in Int.\ Jour.\ of Mod.\ Phys.\ {\bf A} & \cr
\+ &$\q{\zamzam}$
               & A.B.\ Zamolodchikov, Al.B.\ Zamolodchikov & \cr
\+ &           & {\it Conformal Field Theory and Critical Phenomena in
                  Two-Dimensional Systems}  & \cr
\+ &           & Physics Rev.\ Vol.\ 10,4 (1989) p.\ 269  & \cr
\+ & $\q{\gehlen}$
               & G.v.\ Gehlen, V.\ Rittenberg & \cr
\+ &           & {\it Operator Content of the Three-State
                  Potts Quantum Chain} & \cr
\+ &           & J.\ Phys.\ A: Gen.\ {\bf 19} (1986) p.\ L625& \cr
\+ &$\q{\cardy}$
               & J.L.\ Cardy, {\it Effect of Boundary Conditions on
                  the Operator Content of Two-} & \cr
\+ &           & {\it Dimensional Conformally Invariant Theories} & \cr
\+ &           & Nucl.\ Phys.\ {\bf B275} (1986) p.\ 200 & \cr
\+ & $\q{\schuetz}$
               & G.v.\ Gehlen, V.\ Rittenberg, G.\ Sch\"utz & \cr
\+ &           & {\it Operator Content of $n$-State Quantum Chains in the
                  $c=1$ Region} & \cr
\+ &           & J.\ Phys.\ A: Gen.\ {\bf 21} (1988) p.\ 2805& \cr
\+ &$\q{\cardyA}$
               & J.L.\ Cardy, {\it Conformal Invariance and Surface
                  Critical Behaviour} & \cr
\+ &           & Nucl.\ Phys.\ {\bf B240} (1984) p.\ 514 & \cr
\+ & $\q{\gehlenA}$
               & G.v.\ Gehlen, V.\ Rittenberg & \cr
\+ &           & {\it The Spectra of Quantum Chains with Free
                  Boundary Conditions} & \cr
\+ &           & J.\ Phys.\ A: Gen.\ {\bf 19} (1986) p.\ L631 & \cr
\+ & $\q{\henkel}$
               & M.\ Henkel, G.\ Sch\"utz & \cr
\+ &           & {\it Finite-Lattice Extrapolation Algorithms},
                  J.\ Phys.\ A: Gen.\ {\bf 21} (1988) p.\ 2617 & \cr
\+ & $\q{\baakeB}$
               & M.\ Baake, G.v.\ Gehlen, V.\ Rittenberg & \cr
\+ &           & {\it Operator Content of the Ashkin-Teller Quantum Chain --
                   Superconformal and} & \cr
\+ &           & {\it Zamolodchikov-Fateev Invariance: II.\ Boundary Conditions
                   Compatible with} & \cr
\+ &           & {\it the Torus},
                   J.\ Phys.\ A: Gen.\ {\bf 20} (1987) p.\ L487 & \cr
\+ & $\q{\kohmoto}$
               & M.\ Kohmoto, M.\ den Nijs, L.P.\ Kadanoff & \cr
\+ &           & {\it Hamiltonian Studies of the $d=2$ Ashkin-Teller Model}&\cr
\+ &           & Phys.\ Rev.\ {\bf B24} (1981) p.\ 5229 & \cr
\+ & $\q{\baakeA}$
               & M.\ Baake, G.v.\ Gehlen, V.\ Rittenberg & \cr
\+ &           & {\it Operator Content of the Ashkin-Teller Quantum Chain --
                   Superconformal} & \cr
\+ &           & {\it and Zamolodchikov-Fateev Invariance: I.\ Free Boundary
                   Conditions} & \cr
\+ &           & J.\ Phys.\ A: Gen.\ {\bf 20} (1987) p.\ L479 & \cr
\+ & $\q{\alcaraz}$
               & F.C.\ Alcaraz, A.L.\ Santos, {\it Conservation Laws for Z(N)
                   Symmetric Quantum} & \cr
\+ &           & {\it Spin Models and Their Exact Ground State Energies} & \cr
\+ &           & Nucl.\ Phys.\ {\bf B275} (1986) p.\ 436 & \cr
\+ & $\q{\gehlenunpub}$
               & G.v.\ Gehlen, unpublished results from 1986 & \cr
\+ &$\q{\gepner}$
               & D.\ Gepner, Z.\ Qiu & \cr
\+ &           & {\it Modular Invariant Partition Functions for
                   Parafermionic Field Theories} & \cr
\+ &           & Nucl.\ Phys.\ {\bf B285} (1987) p.\ 423 & \cr
\+ &$\q{\saleur}$
               & B.\ Duplantier, H.\ Saleur & \cr
\+ &           & {\it Exact Critical Properties of Two-Dimensional Dense
                    Self-Avoiding Walks} & \cr
\+ &           & Nucl.\ Phys.\ {\bf B290} (1987) p.\ 291 & \cr
\+ &$\q{\duplantier}$
               & B.\ Duplantier, K.\ Kostov & \cr
\+ &           & {\it Geometrical Critical Phenomena on a Random Surface
                    of Arbitrary Genus} & \cr
\+ &           & Nucl.\ Phys.\ {\bf B340} (1990) p.\ 491 & \cr
\vfill
\end